\documentclass[prc,fleqn,twocolumn,superscriptaddress,showpacs,preprintnumbers,amsmath,amssymb]{revtex4}

\usepackage{graphicx}
\usepackage{dcolumn}
\usepackage{bm}
\usepackage{color}

\begin{document}


\title{Probing the density dependence of symmetry energy via multifragmentation at sub-saturation densities\\}

\author{Sanjeev Kumar}

\author{Y. G. Ma\footnote{Author to whom all correspondence should be addressed. Email:
ygma@sinap.ac.cn}}
\affiliation{Shanghai Institute of Applied Physics, Chinese
Academy of Sciences, Shanghai 201800, China}
\author{G. Q. Zhang}

\author{C. L. Zhou}
\affiliation{Shanghai Institute of Applied Physics, Chinese
Academy of Sciences, Shanghai 201800, China} \affiliation{Graduate
School of the Chinese Academy of Sciences, Beijing 100080, China}

\date{\today}

\begin{abstract}
Symmetry energy for asymmetric nuclear matter at sub-saturation
densities was investigated in the framework of an
isospin-dependent quantum molecular dynamics model. Single ratio
of neutrons and protons  is compared, for the first time, with the
experimental data of Famiano {\it et al.} We have also performed
the comparison for double ratio with experimental as well as
different theoretical results of BUU97, IBUU04, BNV and ImQMD
models. It is found that the double ratio predicts the softness of
symmetry energy, which is little underestimated in single ratio.
Further, the study of single ratio is extended for different kind
of fragments, while, double ratio is for different neutron-rich
isotopes of Sn.
\end{abstract}

\maketitle

\section{Introduction}
One of the most important challenges in heavy-ion physics is the
determination of the isospin dependence of nuclear equation of
state (NEOS), which plays very important role at low energy
phenomena like nuclear structure, nuclear astrophysics
\cite{Latt00}, fusion, cluster radioactivity \cite{Amor09} etc;
intermediate energy phenomena like multifragmentation, stopping,
flow \cite{Toro08,
Ma-flow,Kuma10,Li08,Xu00,Colo,Nato,Yell,Fu09,MaCW,
Fami06,Hils87,Li97,Li06,Wolt07,Colo08,Zhan08, Tsan11,Tsan04,Liu03,
Bara02, Ruso11,Li00, Gaut11} etc.; and at last high energy
phenomena like pion and kaon production \cite{Li04,Hart08} etc.
The symmetry energy is found to be the prominent candidate to
study the isospin dependence of NEOS. In the past years, many
studies are performed on the density dependence of symmetry energy
at sub-saturation densities by using isotopic scaling
\cite{Xu00,Colo,Nato,Yell,Fu09}, isobaric ratio \cite{MaCW},
single and double ratios \cite{Fami06,Hils87,Li97,Li06,
Wolt07,Zhan08,Tsan11}, isospin diffusion \cite{Li08,Tsan04},
isospin distillation/fractionation \cite{Li08,Liu03}, and isospin
migration/drift \cite{Toro08,Li08,Bara02} etc. Apart from these,
transverse and elliptic flow of neutrons and protons are also
considered as a good candidate to emphasize on the importance of
density dependence of symmetry energy \cite{Ruso11,Li00,Gaut11}.
Even with the help of these studies, the exact determination of
symmetry energy is still under the way.

In present work, we only want to address the effect of symmetry
energy on kinetic energy spectra of nucleons as well as the
neutrons to protons ratio parameters. The later one was considered
to be the ever first prominent candidate to extract the density
dependence of symmetry energy.

Before we discuss the contents of the present work, let us have
some highlights of the single and double ratio in heavy ion
collisions. The single ratio study in heavy-ion collisions have
already been done by different experimental and theoretical groups
\cite{Fami06,Hils87, Tsan11}. In the experiments, near Fermi
energy, Hilscher {\it et al.} \cite{Hils87} found that single
ratio of pre-equilibrium nucleons is consistently higher than that
of projectile-target system and it can not be explained by the
Coulomb effects alone. Another experimental observation is the
ratio of free neutrons and protons from two isotopic systems at 26
MeV/nucleon. A lot of interesting observations are made from the
data. Schroder {\it et al.} \cite{Li} also systematically studied
the spectra of pre-equilibrium neutrons and protons in both
isospin symmetric and asymmetric systems. Recently, at NSCL/MSU
Famiano {\it et al.} \cite{Fami06} measured the single and double
ratios of free neutrons to protons for $^{112}$Sn + $^{112}$Sn and
$^{124}$Sn + $^{124}$Sn at 50 MeV/nucleon. The results of double
ratio of the above data have also been reproduced by different
theoretical models, such as BUU97 \cite{Li97}, IBUU04 \cite{Li06},
BNV \cite{Wolt07} and ImQMD \cite{Zhan08}. Even, there are a lot
of uncertainties in the determination of symmetry energy in term
of different parameters like cross-section, symmetry energy
coefficient, impact parameter and method of clusterization etc.

However, no study exists in the literature where the comparison of
single ratio of neutrons to protons is performed with the
experimental data. One step ahead, the single ratio for the
fragments is still poorly known in the literature. A few studies
existed from the BNV and IBUU04 calculations for the single ratio
using the intermediate mass fragments (IMF's), which is only
limited for the small range of the kinetic energy
\cite{Li08,Colo08}, however, isospin distillation/fractionation is
studied up to higher kinetic energy by Li {\it et al.}
\cite{Li07}. In extension of single ratio to double ratio, no one
has tried to compare the double ratio findings for experiments and
theories at one place to see which one is the most appropriate
model and symmetry energy form. It is also absent from the
literature that what is the affect on the double ratio if we
consider the series of isotopes with different isospin contents?
From all these gaps, it seems interesting to perform study on the
single and double ratios simultaneously.

In this paper, we focus on the comparative study of single and
double ratios of neutrons to protons with the experimental data of
the MSU/NSCL collaborations \cite{Fami06}. Moreover, in addition
to the comparison with the experimental data, our results of IQMD
model (initially developed by Hartnack {\it et al.} \cite{Hart98})
are also compared with  other studies of BUU97, IBUU04, BNV and
ImQMD models. The study of single ratio is extended for different
kinds of fragments up to higher kinetic energy, while, the double
ratio is investigated for different neutron-rich systems having
different isospin content.

The article is organized as follow: we discuss the model briefly
in Sec. II. Our results and discussions are given in Sec. III and we
summarize the results in Sec. IV.

\section{Formalism: Isospin dependent Quantum Molecular Dynamics Model (IQMD)}
\label{IQMD}

In IQMD model \cite{Kuma10,Gaut11,Hart98}, nucleons are
represented by the wave packets, just like the QMD model
\cite{Aich91}. These wave packets of the target and projectile
interact by the full Skyrme potential energy, which is represented
by $U$ and is given as:
\begin{equation}
U~=~U_{\rho}+U_{Coul}.
\end{equation}
Here $U_{Coul}$ is the Coulomb energy, and  $U_{\rho}$ is originated from the density dependence of the nucleon optical potential and
is given as:
\begin{equation}
U_{\rho}~=~\frac{\alpha}{2}\frac{\rho^2}{\rho_0}~+~\frac{\beta}{\gamma+1}\frac{\rho^{\gamma+1}}{\rho_0^{\gamma}}+\frac{C_{s,p}}{2}
\left(\frac{\rho}{\rho_0}\right)^{\gamma_i}\delta^2\rho,
\label{equation2}
\end{equation}
where $\delta~=~\left(\frac{\rho_n-\rho_p}{\rho_n+\rho_p}\right)$;
$\rho~=~\rho_n+\rho_p$, $\rho_n$ and $\rho_p$ are the neutron and
proton densities, respectively. The densities $\rho$, $\rho_n$ and
$\rho_p$ has the dimensions of $fm^{-3}$.

First two of the three parameters of the Eq.~\ref{equation2}
($\alpha$ and $\beta$) are determined by demanding that at normal
nuclear matter densities, the binding energy should be equal to 16
MeV  and the total energy should have minimum at $\rho_0$. The
third parameter $\gamma$ is usually treated as a free parameter.
Its value is given in term of the compressibility:
\begin{equation}
\kappa~=~
9\rho^{2}\frac{\partial^{2}}{\partial\rho^{2}}\left(\frac{E}{A}\right)~~\cdot
\end{equation}
The different values of compressibility give rise to a soft and a
hard equation of state. The soft equation of state is employed in
the present study with the parameters $\alpha$~=~-356~MeV,
$\beta$~=~303~MeV and $\gamma~$=~7/6 corresponding to isoscalar
compressibility of $\kappa~$=~200~MeV. In the calculations, we use
the isospin dependent in-medium cross section in the collision
term and the Pauli blocking effects just like QMD model
\cite{Aich91}.  The third term in the Eq.~\ref{equation2} is the
symmetry potential energy for a finite nuclear matter. The
symmetry energy per nucleon employed in the simulation is the sum
of the kinetic and potential term. So, the total symmetry energy
is given as:
\begin{equation}
E_{Sym}(\rho)~=~\frac{C_{s,k}}{2}\left(\frac{\rho}{\rho_0}\right)^{2/3}~+~\frac{C_{s,p}}{2}\left(\frac{\rho}{\rho_0}\right)^{\gamma_i}
,
\end{equation}
where, $C_{s,k}$~=~25 MeV from the Fermi Dirac distribution, which
is well explained in Ref.~\cite{thesis}, is known as symmetry
kinetic energy coefficient, while, $C_{s,p}$~=~35.19~MeV is
parametrized on the basis of the experimental value of the
symmetry energy, is known as symmetry potential energy
coefficient. On the basis of $\gamma_i$ value, symmetry energy is
divided into two types with $\gamma_i~=~0.5$ and $\gamma_i~=~1.5$,
corresponds to the soft and stiff symmetry energies, respectively.
Note that the $\gamma$ used in the determination of equation of
state and $\gamma_i$ used in the determination of symmetry energy
are the different parameters. The interesting feature of symmetry
energy is that, its value increases with decreasing $\gamma_i$ at
sub-saturation densities, while, opposite is true at
supra-saturation densities. In other words, the soft symmetry
energy is more pronounced at sub-saturation densities, while the
stiff symmetry energy at supra-saturation densities.

The cluster yields are calculated by means of the  coalescence
model, in which particles with relative momentum smaller than
$P_{Fermi}$ and relative distance smaller than $R_{0}$ are
coalesced into a cluster. The value of $R_{0}$ and $P_{Fermi}$ for
the present work are 3.5 fm and 268 MeV/c, respectively.

\section{Results and Discussion}
\label{Results}

In the present study, we simulate thousands of events for the
isotopes of Sn, namely $^{112}$Sn + $^{112}$Sn, $^{124}$Sn +
$^{124}$Sn and $^{132}$Sn + $^{132}$Sn at incident energy of 50
MeV/nucleon by using the soft and stiff symmetry energy having
$\gamma_i~=~0.5$ and 1.5, respectively. The collision geometry for
the study is from semi-central to semi-peripheral one by keeping
in mind the importance of impact parameter of NSCL/ MSU
collaboration's experimental results. As discussed earlier, the
soft equation of state with an isospin dependent NN cross sections
of
$\sigma_{med}~=~\left(1-0.2\frac{\rho}{\rho_0}\right)\sigma_{free}$
is employed. The single and double ratio is considered as
a point of importance in the present study. The neutrons to
protons ratio is among the first observables that was proposed a
possible sensitive probe for symmetry energy prediction
\cite{Fami06,Hils87, Tsan11}. This ratio is studied for the free
nucleons, light charged particles (LCP's) (having charge number of
 1 and 2) and intermediate mass fragments (IMF's) (having charge
between 3 and $Z_{tot}/6$), where $Z_{tot}$ is the total charge of
the projectile and target under study. The single ratio is just
the ratio of neutrons to protons and is represented in the study
by the $R_{N/Z}$, while double ratio is the ratio of the single
ratios of any two isotopes of the Sn. In order to study the
systematics of the isospin effect, the single ratio of the isotope
with more number of neutrons  is always mentioned in the
numerator, when double ratio is calculated. Mathematically, the
double ratio is represented by $DR_{N/Z}$ and is given as :
\begin{equation}
DR_{N/Z}~=~\frac{R_{N/Z}^{neutron~rich}}{R_{N/Z}^{neutron~weak}}.
\end{equation}

\subsection{Kinetic energy spectra}
In order to go in detail in the results from all above  ratios,
let us understand the kinetic energy spectra of protons and
neutrons for all type of fragments in the center of mass frame.
The spectra of free protons and neutrons are very important
experimental observables that can provide useful information about
the particle production mechanism and reaction dynamics.

\begin{figure}
\includegraphics[width=85mm]{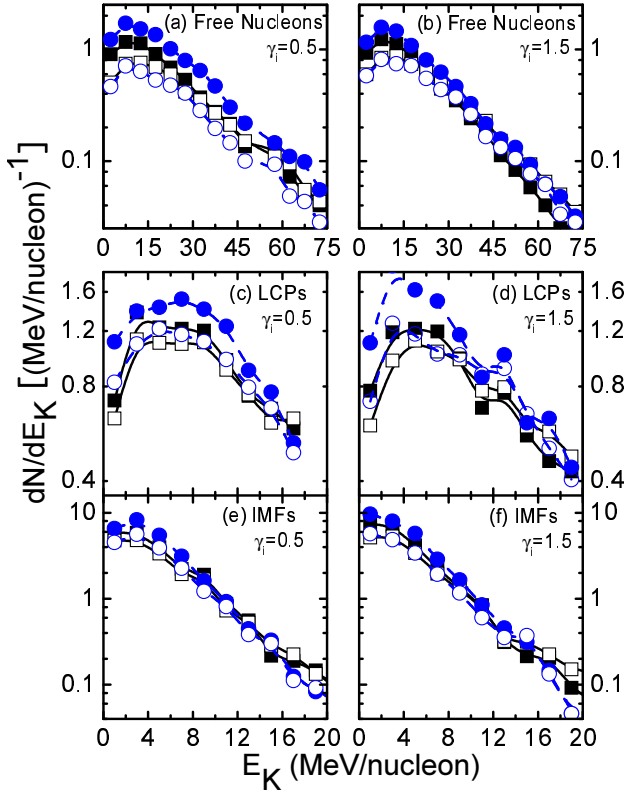}
\caption{\label{fig:1}(Color online)  The kinetic energy spectra
in the center of mass system for neutrons (solid symbols) and
protons (open symbols) from the free nucleons (a,b), LCP's (c,d),
and IMF's (e,f) at semi-central geometry ( b = 2 fm ) of
$^{132}$Sn + $^{132}$Sn (blue circles) and $^{112}$Sn + $^{112}$Sn
(black squares) collisions at E = 50 MeV/nucleon by using the soft
(left) and stiff (right) symmetry energy, respectively.}
\end{figure}

\begin{figure}
\includegraphics[width=85mm]{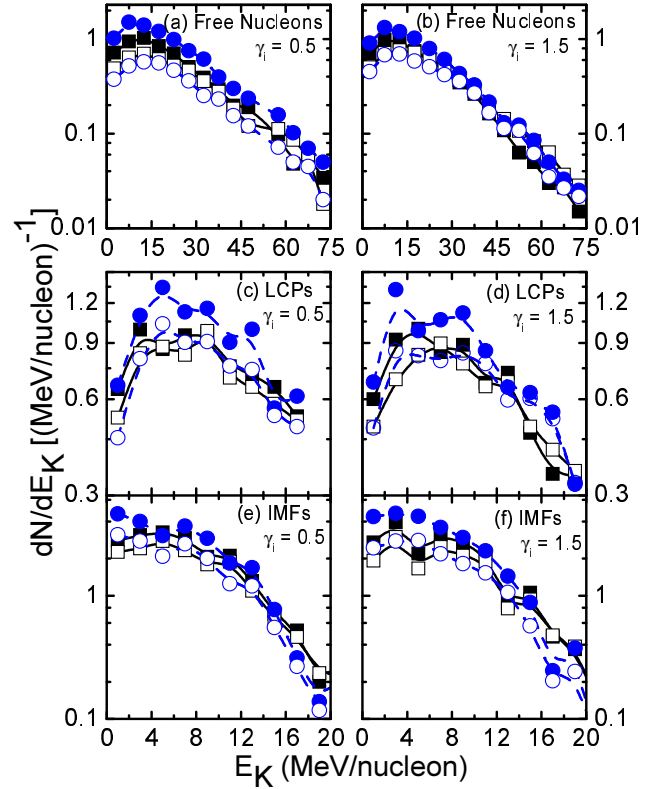}
\caption{\label{fig:2}(Color online)  Same as in Fig. \ref{fig:1},
but for the semi-peripheral geometry (b= 6 fm).}
\end{figure}

Fig. \ref{fig:1} displays the kinetic energy spectra for protons
and neutrons at incident energy E = 50 MeV/nucleon for
semi-central geometry, while, Fig. \ref{fig:2} is at
semi-peripheral geometry. The results are for neutron-rich
$^{132}$Sn~+~$^{132}$Sn and neutron-weak system $^{112}$Sn +
$^{112}$Sn by using the soft and stiff symmetry energy,
respectively. The left and right panels in both figures are  with
the soft and stiff symmetry energy, while top, middle and bottom
panels are for the free nucleons, and bound nucleons inside LCP's
and IMF's, respectively.

It is clear from the figure that the production of neutrons is
more favorable for neutron-rich system \cite{Ma-99}. It is also
true for all types of fragments as well as for the soft and stiff
symmetry energy. This is due to the reason that in more
neutron-rich system, the symmetry energy is more repulsive
(attractive) for neutrons (protons) and hence more neutrons can be
produced. On the second hand,  the difference between yield or
content of neutrons and protons decreases with the increasing of
the kinetic energy. This is due to the Coulomb repulsion, which
shifts the protons from low to high kinetic energy. The behavior
is the same for all types of fragments. Interestingly, more
neutrons can be produced with the soft symmetry energy for free
particles as compared to the stiff one. The opposite is true for
the LCP's and IMF's up to a certain kinetic energy  and after that
the same trend is observed just like for the free particles. It is
an interesting phenomenon and unfortunately no one has noticed
this one. This is due to the reason that the Coulomb effects are
stronger {\color{red} inside} LCP's and IMF's as compared to free
particles. However, at a sufficient high kinetic energy, symmetry
energy dominates over the Coulomb interactions and the behavior
becomes just similar to that of free nucleons. Although, this
intersection between soft and stiff symmetry energy for the
fragments is not so clearly observed from here, so, we have
extended the study with single ratio $R_{N/Z}$ in the Fig.
\ref{fig:3}.

Let us move to the Fig. \ref{fig:2}, which is displayed at
semi-peripheral geometry. Almost, the same spectra are observed at
semi-peripheral geometry, except for some exceptions. Once again,
interesting point is that yield of free neutrons at high kinetic
energy for semi-central geometry (Fig. \ref{fig:1}) is higher in
comparison with the semi-peripheral one (Fig. \ref{fig:2}). As we
already knew that the symmetry energy/potential has two important
functions: firstly, it tends to unbound more neutrons and
secondly, it makes the neutrons more energetic than protons. Due
to this, most of the finally observed neutrons are unbounded in
the very early stage of the reaction as a result of the
nucleon-nucleon (NN) collisions at semi-central collisions. Now,
symmetry energy at pre-equilibrium time is just shifting the more
neutrons towards the high kinetic energy. On the other hand, at
semi-peripheral geometry, the emission of neutrons also depend on
the symmetry potential/energy due to the relative lack of the NN
collisions. The symmetry energy makes the neutrons unbound, but at
relatively low kinetic energy. That is why, isospin effects are
more pronounced at low kinetic energy for peripheral collisions
and at high kinetic energy for central collisions. These results
are also consistent with those in Ref.~\cite{Li08}. When one moves
from LCP's to IMF's, the content of neutrons for neutron-rich
system is lower than the neutron-weak system at high kinetic
energies for semi-central as well as semi-peripheral geometries.
This is due to the reason that at high kinetic energy the yield of
free nucleons is higher as compared to fragments and it will
result in more production of free neutrons for neutron-rich system
as compared to the fragments.

\subsection{Single ratio}

In order to make sure about the above discussion from Fig.
\ref{fig:1} and Fig. \ref{fig:2}, it is interesting to investigate
the single ratio ($R_{N/Z}$) of neutrons to protons for free
nucleons, LCP's and IMF's, which is shown in Fig. \ref{fig:3} for
neutron-rich system $^{132}$Sn~+~$^{132}$Sn and neutron-weak
system $^{112}$Sn + $^{112}$Sn by using the soft and stiff
symmetry energy. The left and right panels are at semi-central and
semi-peripheral geometries, respectively. As is expected from Fig.
\ref{fig:1} and Fig. \ref{fig:2}, Fig. \ref{fig:3} depicts the
results as follow:
\begin{itemize}

\item{The isospin effects for more neutron-rich system are
stronger and it is consistent with Ref.~\cite{Li08} and with Figs.
\ref{fig:1}, \ref{fig:2}.}

\item{$R_{N/Z}$ decreases with the kinetic energy for all types of
fragments at semi-central as well as semi-peripheral geometries.}

\item{ For free nucleons, the isospin effects are stronger at high
kinetic energy for semi-central geometry, while, the same is true
at low kinetic energy for semi-peripheral geometries. It is also
explained earlier in Ref.~\cite{Li08}.}

\item{The increase in the neutrons to protons ratio for
neutron-rich system  at sufficient high kinetic energy is due to
the repulsive nature of the symmetry energy for neutrons.}

\item{Let us discuss the single ratio for fragments. $R_{N/Z}$ of
IMF's is earlier studied by the Catania group using the BNV
\cite{Colo08} and Texas group using the IBUU04 model \cite{Li08}.
Both models have different approaches for the symmetry energy and
hence the results are little different from each other. In the BNV
results, ratio decreases at low fragment kinetic energy and then
increases at high kinetic energy for neutron-rich system with the
stiff symmetry energy. On the other hand, in the IBUU04
calculations, the ratio is found to decrease with fragment kinetic
energy. However, both the calculations have the same behavior with
the  soft and stiff symmetry energy. But, both groups have limited
their study only to the relative low kinetic energy and were not
able to investigate the cross-over phenomenon of symmetry energy,
which takes place at higher kinetic energies and discussed in
detail in this study.}

\begin{figure}
\vspace{-1.0cm}
\includegraphics[width=85mm]{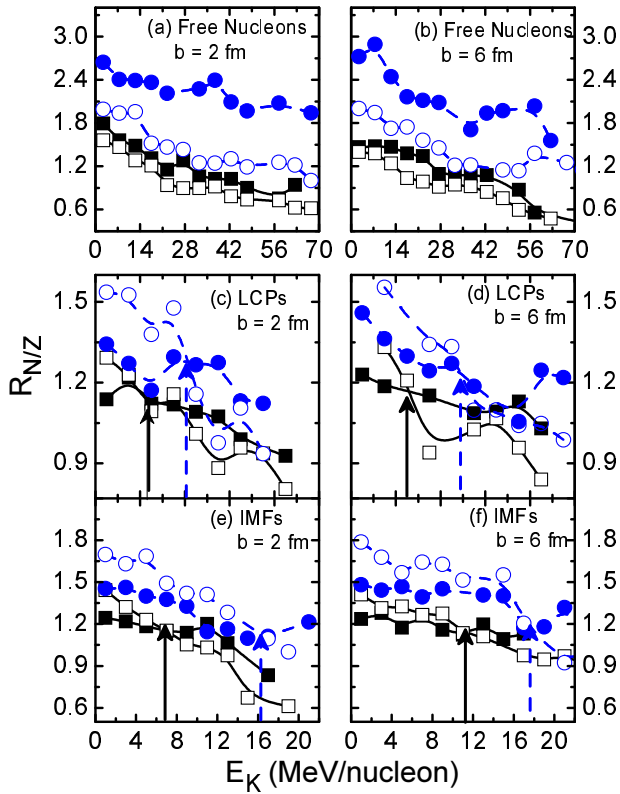}
\caption{\label{fig:3} (Color online) The ratio of neutrons to
protons (a,b) at freeze-out time from free nucleons, LCP's (c,d)
and IMF's (e,f) as a function of kinetic energy at semi-central
(left) and semi-peripheral (right) geometries by using the soft
(solid symbols) and stiff symmetry energy (open symbols). The
vertical lines in the plots of LCP's and IMF's represent the
kinetic energy at the cross-over points of the soft and stiff
symmetry energy. Blue circle represents for $^{132}$Sn +
$^{132}$Sn and black squares for $^{112}$Sn + $^{112}$Sn at E = 50
MeV/nucleon.}
\end{figure}

\item{The large isospin effects are observed with the soft
symmetry energy for free nucleons along whole range of the kinetic
energy \cite{Tsan11}, while cross-over happens for the LCP's and
IMF's at certain kinetic energy. Below the cross-over kinetic
energy, the stiff symmetry energy produces larger neutrons to
protons ratio and after the cross-over, it is true with the soft
symmetry energy and behaves like just for free nucleons. Recently,
Harmann {\it et al.} \cite{ect} displayed the data for single
ratio from the IMF's below the cross-over kinetic energy. This
data (not shown here) is favoring the soft symmetry energy in our
studies with the IQMD, however, with the BNV, there data is well
explained by the stiff symmetry energy. If one sees them
carefully,  we can find that the soft symmetry energy is more soft
and stiff symmetry energy is less stiff in the BNV as compared to
the IQMD and ImQMD models. In other words, the stiff symmetry
energy from the BNV and the soft symmetry energy from the
IQMD/ImQMD lies between the stiff symmetry energy from the
IQMD/ImQMD and the soft symmetry energy from the BNV
calculations}. It means that the data is favoring almost the same
symmetry energy from both the models.

\begin{figure}
\includegraphics[width=80mm]{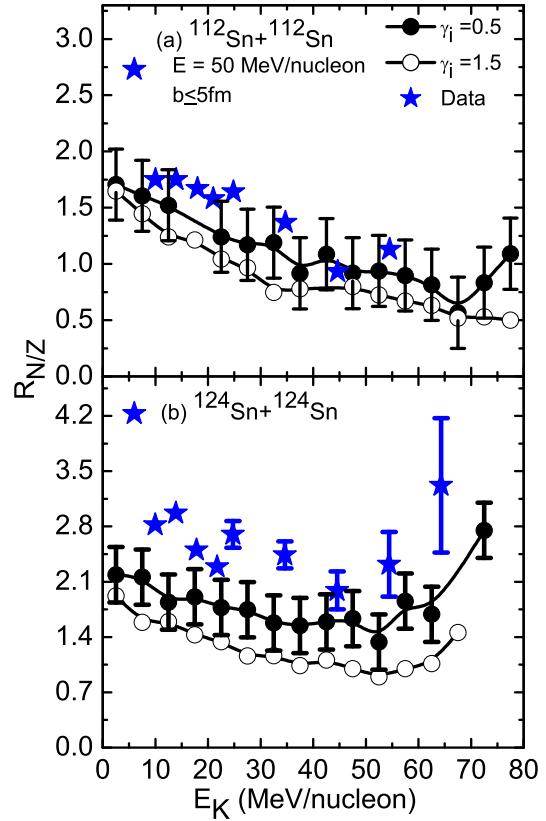}
\caption{\label{fig:4}(Color online)  The comparison of neutrons
to protons ratio from free nucleons, for the systems $^{112}$Sn +
$^{112}$Sn (top) and $^{124}$Sn + $^{124}$Sn (bottom) at E = 50
MeV/nucleon and impact parameter $b\le5$, with the experimental data of
MSU/NSCL collaborations \cite{Fami06}. The filled and  open
circles represent the soft and stiff symmetry energies,
respectively.}
\end{figure}

\item{The cross-over kinetic energy is at higher value for more
neutron-rich system and increases with  the size of the fragments
(i.e. from LCP's to IMF's).}

\item{The cross-over value of the kinetic energy also raises when
one moves from semi-central to semi-peripheral geometries. This
value is more affected for the more neutron-rich system.}

\item{It is also clear from here that gas phase (free nucleons) is
significantly enriched in neutrons relative to the liquid phase or
fragments that are represented by the bounding nuclei. The
phenomenon is known as isospin distillation/fractionation and is
discussed many times in the literature only in term of the free
and bound nucleons \cite{Li08}. More interesting results are
expected for isospin distillation if one tries to study in term of
different kind of fragments.}

\end{itemize}

The theoretical results become more interesting and useful if one
compares the results with experimental data. In the Fig.
\ref{fig:4}, we have, for the first time, compared the results of
single ratio of neutrons to protons from free nucleons for
neutron-weak system $^{112}$Sn~+~$^{112}$Sn (in top panel) and for
neutron-rich system $^{124}$Sn~+~$^{124}$Sn (in bottom panel) at
50 MeV/nucleon with the experimental data \cite{Fami06}. The behavior of
single ratio results for free nucleons is explained in the Fig.
\ref{fig:3}. The conclusion from the figure is that (1) more
$R_{N/Z}$ is observed for more neutron-rich system, which is also
predicted by theoretical predictions,  and (2) $R_{N/Z}$ shows
increment at higher kinetic energy, especially for
$^{124}$Sn~+~$^{124}$Sn. It indicates that the theoretical results
are consistent with the experimental one.

The results are in good agreement with the soft symmetry energy
except at very low and very high kinetic energy. The difference
between soft and stiff symmetry energy results for the
neutron-weak system is almost comparable to the error bar, while,
for neutron-rich system, the difference has a great importance
over the error bar. In other words, the error bar of the
theoretical results with the soft symmetry energy covers the error
bar of the experimental data for both the systems under
consideration. The difference at high kinetic energy between
theoretical and experimental results for neutron-rich system is
due to the large uncertainty in the measurement of the $R_{N/Z}$.
 By using the single ratio observable, one can reach at a partial
conclusion that the asymmetric nuclear matter favors the soft
symmetry energy at sub-saturation densities, which is also
consistent with the other findings in the literature
\cite{Toro08,Li97,Li06, Wolt07,Zhan08,Tsan11, Ruso11}.

As we have observed, the single ratio mixes the symmetry energy
with Coulomb effects throughout the kinetic energy range. In order
to minimize the Coulomb effects and systematical error, it is
reasonable to study the double neutrons to protons ratio for the
isotopes of the same element. This is also studied in the
literatures with only two isotopes \cite{Li97, Li06, Zhan08,
Tsan11}. No one has tried to investigate the effect of double
ratio on a series of isotopes in the asymmetric nuclear matter so
far.

\subsection{Double ratio}

\begin{figure}
\vspace{-1.5cm}
\includegraphics[width=80mm]{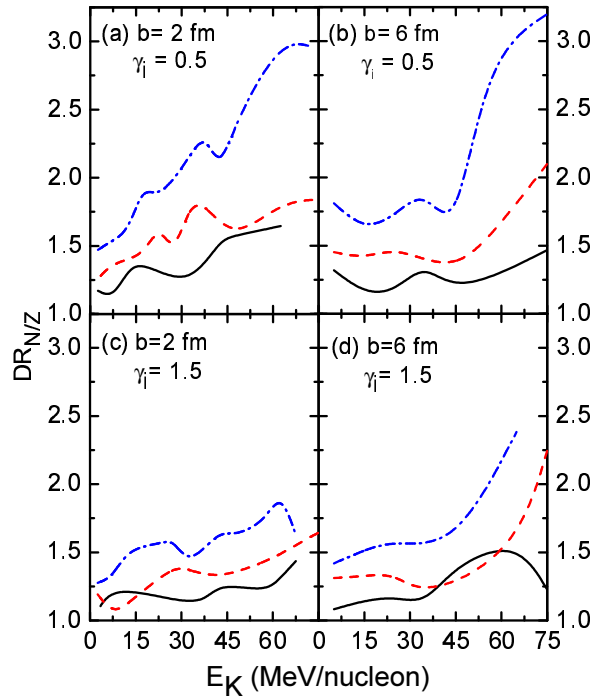}
\caption{\label{fig:5}(Color online) Free neutrons to protons
double ratio, at semi-central (left) and semi-peripheral (right)
geometry with the soft (top) and stiff (bottom) symmetry energy,
as a function of kinetic energy at the incident energy E = 50
MeV/nucleon. The different lines in the figure are the double ratio from
different pairs: Solid line for $^{132}$Sn + $^{132}$Sn and
$^{124}$Sn + $^{124}$Sn; Dashed line for $^{124}$Sn + $^{124}$Sn
and $^{112}$Sn + $^{112}$Sn; Dash-dot line for $^{132}$Sn +
$^{132}$Sn and $^{112}$Sn + $^{112}$Sn.}
\end{figure}

In the present study, we consider reactions between three isotopes
of Sn and observe the relative effect of these isotopes on the
double ratio and symmetry energy. The pairs are as follow:
$^{132}$Sn~+~$^{132}$Sn and $^{124}$Sn~+~$^{124}$Sn,
$^{124}$Sn~+~$^{124}$Sn and $^{112}$Sn~+~$^{112}$Sn,
$^{132}$Sn~+~$^{132}$Sn and $^{112}$Sn~+~$^{112}$Sn. The three
pairs having the difference of 8, 12 and 20 neutrons,
respectively. The universal behavior for double ratio is observed
with the kinetic energy, i.e. with the increasing of kinetic
energy, the double ratio is found to increase for all the three
sets of isotopes which we have plotted in the Fig. \ref{fig:5}.
The increase in the double ratio is due to the effect that now
energetic nucleons are more affected by the symmetry potential,
which are already suppressed by the Coulomb repulsion in the
single ratio results. The effect of symmetry energy on the double
ratio is just like the single ratio, i.e. larger value with the
soft symmetry energy as compared to the stiff one. Moreover,
double ratio goes on increasing  with the increase of the neutron
difference between the pairs discussed above, or in other words,
the double ratio from free nucleons goes on increasing with the
initial-state double ratio of the systems from three different
pairs of isotopes of Sn.

This increase is due to the effect that the more the number of
neutrons, the more repulsive the symmetry energy for them. The
Coulomb effects are already cancelled by taking the double ratio.
Hence, the results are just like that as expected. The double
ratio is found to be weakly sensitive towards the collision
geometry. However, a little increase is observed at
semi-peripheral geometry compared to semi-central one at high
kinetic energy. This is true with the stiff as well as the soft
symmetry energy.

\begin{figure}
\includegraphics[width=80mm]{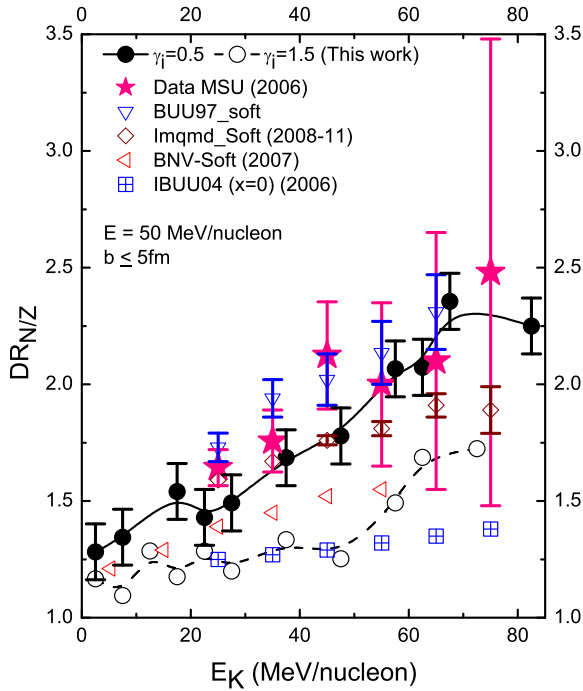}
\caption{\label{fig:6}(Color online) The comparison of free
neutrons to protons double ratio at E = 50 MeV/nucleon and $b\le5$
with the MSU/NSCL data,  BUU97, IBUU04, BNV, and ImQMD
simulations.}
\end{figure}

The double ratio has been studied many times in last couple of
years by different groups with the help of the
 BUU97, IBUU04, BNV and ImQMD models and
compared with the experimental results. Even so, we are still far
away from the exact conclusion about the symmetry energy form. We
have, along with all the possible results in the literature,
compared the double ratio with the IQMD model in Fig. \ref{fig:6}.
Let us start with very first comparison of the BUU97 \cite{Li97}.
The results were very close to the experimental one, but, the
reaction conditions were different. Firstly, in the BUU97
calculations, the incident energy was 40 MeV/nucleon, not 50
MeV/nucleon, just like the experimental one. Secondly, data set is
only for the transverse emission, while in the BUU97 calculations,
the nucleons used are emitted in all the directions. Move one step
ahead to the IBUU04 results \cite{Li06}, where the symmetry energy
is introduced with the help of momentum dependent interactions,
the results are very far from from the experimental data. The same
is true for the BNV calculations performed by the Catania group in
2007 \cite{Wolt07}. The most closeness between the data and the
calculation is observed by the ImQMD model in 2009 \cite{Zhan08}.
They found that the results with $\gamma_i~=~0.75$ are best fit
with the experimental data for impact parameter $b~\le~2$ fm. In
the present study, we have performed simulations for $b~\le~5$ fm
 and for the angular cuts, as mentioned in the experiments, with
the soft and stiff symmetry energy and displayed the theoretical
results over the whole range of the kinetic energy. The Fig.
\ref{fig:6} clearly indicates that our results are very close to
experimental data.

If we see the comparison of theoretical and experimental results
from single (Fig. \ref{fig:4} ) and double ratio  (Fig.
\ref{fig:6} ) results, it seems that single ratio results require
$\gamma_i<0.5$ to explain the data, and while the data is well
explained by the $\gamma_i=0.5$ for double ratio. This is due to
the reason that single ratio suffers the effect from the Coulomb
interactions in addition to symmetry energy. As our main purpose
is to extract the symmetry energy, where double ratio can act as a
better candidate rather than the single ratio. In conclusion, the
results of double ratio can be very well explained by the soft
symmetry energy with $\gamma_i~=~0.5$ in comparison with single
ratio, where the data is little underestimated by the theoretical
predictions.

\section{Summary}
\label{con}

In summary, we have performed a detailed analysis for  kinetic
energy spectra of free nucleons and bound nucleons inside
fragments as well as the ratio parameters for the three reaction
channels of Sn-isotopes at E = 50 MeV/nucleon via
multifragmentation. The kinetic energy spectra of protons and
neutrons from free nucleons and all type of fragments in the
center of mass frame shows that the content of neutrons is more
favorable for neutron-rich system since the symmetry energy
becomes more repulsive (attractive) for neutrons (protons) and
hence more neutrons can be produced. In addition, the difference
between yields or contents of neutrons and protons decreases with
the increasing of the kinetic energy, which can be explained by
the Coulomb repulsion shifts, making the protons from low to high
kinetic energy. Interestingly, more  free neutrons can be produced
with the soft symmetry energy as compared to the stiff one.

From the single ratio of free neutrons to protons, it decreases
with the kinetic energy for all types of fragments at semi-central
as well as semi-peripheral geometries. However, the increase of
the ratio from free nucleons for neutron-rich system is observed
at sufficient high kinetic energy, this can be explained by the
repulsive nature of the symmetry energy for neutrons. For single
ratios of LCP's and IMF's, we noticed a transition
at certain kinetic energy
between the soft and stiff symmetry energy, while no transition
for the free nucleons or gas phase. Below the cross-over kinetic
energy, the stiff symmetry energy produces larger ratio of
neutrons to protons and after the cross-over, it is true with the
soft symmetry energy and behaves like just for free nucleons. This
transition is also found to be strongly dependent on the isospin
of the colliding partners, size of the fragment and weakly
dependent on the collision geometry. Moreover, isospin
distillation is also observed when one moves from the gas phase to
liquid phase. It is further more interesting to study the isospin
distillation in term of different kind of the fragments as
compared to consider bound fragments as a single liquid phase.

The comparison of the theoretical results of single and double
ratios with the experimental data  emphasizes on the softness of
the symmetry energy at sub-saturation densities, which is yet
uncertain at the supra-saturation densities. Although, the single
ratio study  underestimates the data a little as compared to
double ratio for the same stiffness of symmetry energy
($\gamma_i=0.5$), which reflects the double ratio is a relative
good candidate for density dependence of symmetry energy at
sub-saturation densities because of the cancel of Coulomb effect
between two systems. Of course, the magnitude of double ratio of
neutrons and protons from free nucleons strongly depends on the
initial double ratio of the systems. It gives us an indication
that it is better to study the isospin physics with a pair of
$^{132}$Sn + $^{132}$Sn and $^{112}$Sn + $^{112}$Sn.

\begin{acknowledgments}
This work is supported in part by the Chinese Academy of Sciences
Support Program for young international scientists under the Grant
No. 2010Y2JB02, the 973-Program under contract No. 2007CB815004,
and  the National Science Foundation  of China under contract No.s
11035009, 11005140, 10979074,  the Knowledge Innovation Project of
the Chinese Academy of Sciences under Grant No. KJCX2-EW-N01.
\end{acknowledgments}

\end{document}